\documentclass[pre,aps,twocolumn]{revtex4}

\usepackage{graphicx}
\usepackage{amsmath}
\usepackage{amssymb}
\usepackage{ifthen}
\usepackage{booktabs}
\usepackage{color}

\usepackage{graphicx}
\usepackage{dcolumn}
\usepackage{bm}
\usepackage{longtable}
\usepackage{gensymb}

\newcommand{\bb}{\begin{equation}}
\newcommand{\ee}{\end{equation}}
\newcommand{\ba}{\begin{eqnarray*}}
\newcommand{\ea}{\end{eqnarray*}}
\newcommand{\rhor}{\rho({\bf r})}
\newcommand{\dd}{{\rm d}}
\newcommand{\rr}{{\mathbf r}}
\newcommand{\dr}{{\rm d}{\bf r}}

\usepackage{graphicx}
\usepackage{dcolumn}
\usepackage{bm}
\usepackage{longtable}

\begin{document}


\title{Does surface roughness amplify wetting?}

\author{Alexandr \surname{Malijevsk\'y}}
\affiliation{
{Department of Physical Chemistry, Institute of Chemical Technology, Prague, 166 28 Praha 6, Czech Republic}\\
 {Institute of Chemical Process Fundamentals, Academy of Sciences, 16502 Prague 6, Czech Republic}}

\begin{abstract}
Any solid surface is intrinsically rough on the microscopic scale. In this paper, we study the effect of this roughness on the wetting properties of hydrophilic
substrates. Macroscopic arguments,  such as those leading to the well-known Wenzel's law, predict that surface roughness should \emph{amplify} the wetting properties
of such adsorbents. We use a fundamental measure density functional theory (DFT) to demonstrate the opposite effect from roughness for microscopically corrugated
surfaces, i.e., wetting is \emph{hindered}. Based on three independent analyses we show that microscopic surface corrugation increases the wetting temperature or even
makes the surface hydrophobic. Since for macroscopically corrugated surfaces the solid texture does indeed amplify wetting there must exist a crossover between two
length-scale regimes that are distinguished  by opposite response on surface roughening. This demonstrates how deceptive can be affords to extend the thermodynamical
laws beyond their macroscopic territory.
\end{abstract}


\maketitle

\section{Introduction}

It is well known that the geometric or chemical inhomogeneity of a solid surface can dramatically change its adsorption properties \cite{degen, israel, quere}. For
instance, even the most hydrophobic surfaces that are considered smooth on the macroscopic scale, can be characterized by the contact angle of a sessile liquid drop
that typically does not substantially exceed $\theta\approx120\degree$. The equilibrium contact angle is defined by Young's equation \cite{young_p}
 \bb
 \gamma_{\rm sv}=\gamma_{\rm sl}+\gamma\cos\theta\,,\label{young}
 \ee
in terms of the tensions of the solid-vapor, solid-liquid  and liquid-vapor  interfaces. However, if the same material is textured, i.e. the surface is \emph{rough},
the micro- or nanostructure of the surface may induce contact angles close to $180\degree$. This effect has important applications in modern technologies for the
fabrication of devices that utilize super-hydrophobic surfaces. Advances in these technologies are often inspired by natural self-cleaning properties \cite{selfclean}
of certain plant leaves and incest wings where the combination of hydrophobicity and roughness leads to the Lotus effect where dirt particles are adsorbed and removed
by water droplets rolling off these surfaces \cite{koch}. These super-hydrophobic surfaces are replicated in nanotechnologies by grafting polymer chains onto a
substrate \cite{brush} or using nanoimprint lithography \cite{lito}.

Macroscopically, the effect of roughness on substrate wettability is often expressed in terms of the effective (or apparent) contact angle $\theta^*$, which
corresponds to the equilibrium configuration of a macroscopic liquid drop sitting on a rough surface.  One of the most popular relations between Young's and effective
contact angles provides Wenzel's law [9]: assuming that a liquid is in complete contact with a rough surface, i.e., the liquid enters into the grooves of the surface
beneath the drop, simple thermodynamic arguments dictate that:
 \bb
 \cos\theta^*=r\cos\theta\,, \label{wenzel}
 \ee
where $r>1$ is a roughness parameter defined as the ratio of the actual area of the solid surface to the normally projected area.

The most important qualitative conclusion that can be drawn from the Wenzel equation is that surface roughness always amplifies the wetting properties of a given
surface. Therefore, surface roughness makes hydrophobic surfaces (or, more generally, surfaces exhibiting partial drying toward a given fluid) even more hydrophobic,
i.e.,  $\theta^*>\theta>90\degree$, in line with the aforementioned examples of super-hydrophobicity. However, hydrophilic surfaces (or, more generally, surfaces
exhibiting partial wetting toward a given fluid) are predicted to be rendered hydrophilic still more by surface roughness, since $\theta^*<\theta<90\degree$,
according to Eq.~(\ref{wenzel}).



It is considerably more challenging to develop a description of substrate topography effects at the microscopic scale. On the molecular level, the competition between
the fluid-fluid and the fluid-wall interactions, as well as the detailed structure of the wall must be properly considered. While wetting phenomena on structureless
substrates are fairly well understood \cite{dietrich, sullivan, schick, bonn}, a connection between adsorption on microscopically corrugated surfaces and the
underlying intermolecular forces is still largely missing. One obvious reason for the absence of such a description is that in contrast with structureless walls, that
present a one-dimensional problem \cite{tar_ev}, the two- or three-dimensional problem of rough surfaces is much more involved computationally, such that the
literature involving theoretical \cite{bryk, pizio, dutka, rucken, berim, zhou} and simulation \cite{binder, kumar, seveno} studies is rather limited.

In this paper, we study the wetting behavior of rough solid surfaces using a microscopic density functional theory (DFT) \cite{evans79}. In particular, we determine
how the surface roughness affects the wetting properties of a hydrophilic substrate. Our DFT is based on the Tarazona version \cite{tarazona2000, tarazona2002} of the
fundamental measure theory (FMT) \cite{rosenfeld89}, which accurately captures short-range correlations between particles and satisfies exact statistical mechanical
sum rules. The rough substrate is modeled as a semi-infinite planar wall onto which a linear array of tiny pillars is deposited. Translation symmetry along one
Cartesian dimension is assumed and gravity is ignored.
The model is similar to the one considered in Ref. \cite{mal_grooves} for a grooved substrate, but differs in several aspects: i) The corrugation of the wall is much
finer and represents an intrinsic surface roughness  rather than a fabricated structure. Thus, capillary effects inside the grooves between two neighboring pillars
play only a minor role and the dimension of these grooves cannot induce a filling transition. Therefore, the only relevant surface phase transition is the wetting
transition, when the contact angle of a sessile drop vanishes. We particularly wish to determine how the temperature at which the wetting transition occurs depends on
the wall geometry. ii)   In Ref. \cite{mal_grooves}, the dispersion force exerted by the wall was induced by the interaction between solid atoms and fluid atoms via
the attractive part of the Lennard-Jones potential. The repulsive part of the interaction was modeled by coating the outer layer of the solid atoms with a hard wall
of a thickness corresponding to the fluid atom diameter. Such a model somewhat exaggerates the repulsion near the apex of rectangular ridges, which in turn
overestimates the attraction of the interfacial potential that binds the wetting film to the substrate due to the edge-shaped geometry of the substrate
\cite{parry_apex, mal_apex}. In this study, the interaction between the wall and fluid atoms is described by the full Lennard-Jones potential avoiding the need to
introduce an extra hard-wall repulsion, which makes a description of the fluid structure near the edges more realistic. iii) In our previous studies, we described the
strongly oscillating structure of the fluid that is induced by strong confinement using the original Rosenfeld's FMT \cite{rosenfeld89}. In the present study, we
employ a more sophisticated version of the FMT involving tensorial weighted densities that ensure the correct performance of the resulting functional in reduced
dimensions, which is important for strongly confining geometries. This functional is probably the most satisfactory description of packing effects within the
currently available approximations of the intrinsic free energy for nonuniform fluids.

The primary result of this study is that,  contrary to macroscopic predictions, the microscopic surface roughness always deteriorates the substrate wettability. We
demonstrate this phenomenon using three different numerical DFT analyzes: 1) We use a standard grand-canonical ensemble DFT to determine the wetting temperature $T_w$
and the contact angle temperature dependence for $T<T_w$. 2)  We use a constrained DFT (cDFT) where we fix the average number of particles per unit length to
determine the equilibrium shape of the liquid-vapor interface.
3) Finally, we contrast the wetting behavior of smooth and rough surfaces by considering a substrate which is an assemble of both types of surfaces.  Using both DFT
and cDFT we demonstrate that wetting of the rough sections is disfavored until a relatively high temperature that correspond to the wetting temperature of the rough
section, when the entire surface becomes completely wet.

 Before concluding this section we want to emphasize that the intention of this work is \emph{not} to scrutinize the limit of validity of Wenzel's relation Eq.~
(\ref{wenzel}) on the macroscopic scale, as has already been done in numerous studies (see, e.g., Refs.~\cite{quere, wolansky, bico, ishino, quere2}). Instead, we
want to highlight that \emph{any} approach based solely on phenomenological arguments fails to account for even a qualitative description of wetting for
microscopically rough surfaces. This failure can particularly well be illustrated on the simple Eq.~(\ref{wenzel}) but it is no specific feature of the Wenzel's
relation. We will come back to this point in the final section.



The remainder of the paper is organized as follows. In the following section, we define the microscopic model and  formulate the DFT and cDFT treatments. The
numerical results are presented in section III: we first describe the wetting properties of a smooth wall and then contrast these results  with the wetting properties
of corrugated substrates. The results are summarized in section IV and discussed in the concluding section V.

\section{Density functional theory}

\subsection{Grand canonical ensemble DFT}

\begin{figure}
\includegraphics[width=0.5\textwidth]{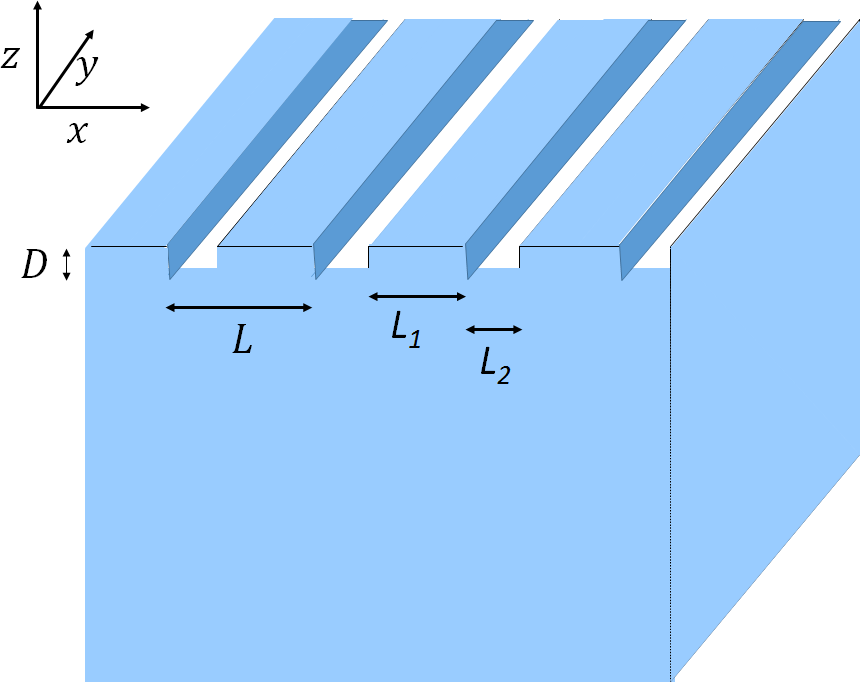}
\caption{Schematic picture of our model of a rough surface. The pillars of height $D$ and width $L_1$ are deposited on a semi-infinite slab with a periodicity $L$.
The model is assumed to be translation invariant along the $y$-axis.} \label{sketch}
\end{figure}

Within the classical density functional theory formulated by Evans \cite{evans79}, the equilibrium density profile is determined by minimizing the grand potential
functional:
 \bb
 \Omega[\rho]={\cal F}[\rho]+\int\dd\rr\rhor[V(\rr)-\mu]\,,\label{om}
 \ee
where $\mu$ is the chemical potential and $V(\rr)$ is the external potential. The intrinsic free energy functional ${\cal F}[\rho]$ can be divided into an exact ideal
gas contribution and an excess part:
  \bb
  {\cal F}[\rho]=\frac{1}{\beta}\int\dr\rho(\rr)\left[\ln(\rhor\Lambda^3)-1\right]+{\cal F}_{\rm ex}[\rho]\,,
  \ee
where $\Lambda$ is the thermal de Broglie wavelength, which can be set to unity and $\beta=1/k_BT$ is the inverse temperature. Following a standard perturbation
approach, the excess term is further split into  hard-sphere and attractive contributions where the latter is treated in a simple mean-field fashion. Thus we write
  \bb
  {\cal F}_{\rm ex}[\rho]={\cal F}_{\rm hs}[\rho]+\frac{1}{2}\int\int\dd\rr\dd\rr'\rhor\rho(\rr')u_{\rm a}(|\rr-\rr'|)\,, \label{f}
  \ee
where  $u_{\rm a}(r)$ is the attractive part of the fluid-fluid interaction potential.

In our model, the fluid atoms are assumed to interact with each other via a truncated (but non-shifted) Lennard-Jones-like potential
 \bb
 u_{\rm a}(r)=\left\{\begin{array}{cc}
 0\,;&r<\sigma\,,\\
-4\varepsilon\left(\frac{\sigma}{r}\right)^6\,;& \sigma<r<r_c\,,\\
0\,;&r>r_c\,.
\end{array}\right.\label{ua}
 \ee
which is cut-off at $r_c=2.5\,\sigma$, where $\sigma$ is the hard-sphere diameter. Hereafter, we will use the parameters $\sigma$ and $\varepsilon$ as the length
and energy units.

The hard-sphere part of the excess free energy is approximated by the fundamental measure theory (FMT) functional \cite{rosenfeld89}:
 \bb
{\cal F}_{\rm hs}[\rho]=\frac{1}{\beta}\int\dd\rr\,\Phi(\{n_\alpha\})\,.\label{fmt}
 \ee
 Within the original Rosenfeld's approach, the weighted densities $n_\alpha$ consist of four scalar and two
vector functions, which are given by convolutions of the density profile and the corresponding weight function:
 \bb
 n_\alpha(\rr)=\int\dr'\rho(\rr')w_\alpha(\rr-\rr')\;\;\alpha=\{0,1,2,3,v1,v2\}\\ \label{n_alpha}
 \ee
where $w_3(\rr)=\Theta(R-|\rr|)$, $w_2(\rr)=\delta(R-|\rr|)$, $w_1(\rr)=w_2(\rr)/4\pi R$, $w_0(\rr)=w_2(\rr)/4\pi R^2$, $w_{v2}(\rr)=\frac{\rr}{R}\delta(R-|\rr|)$,
and $w_{v1}(\rr)=w_{v2}(\rr)/4\pi R$. Here,  $\Theta(r)$ is the Heaviside function, $\delta(r)$ is the Dirac delta function and the hard-sphere radius was set to
$R=\sigma/2$.

It is well known that the original Rosenfeld's functional and all of the alternative approaches that use the set of weighted densities according to (\ref{n_alpha})
provide an excellent description of short range correlations and satisfy the exact statistical mechanical sum rules and thermodynamic conditions at planar walls and
corners \cite{our_wedge}. However, this class of functionals fails to describe the hard-sphere crystal and can produce spurious divergences for highly packed systems
beyond the planar geometry \cite{mulero}. Here, we use the FMT version proposed by Tarazona \cite{tarazona2000, tarazona2002} in which the set of weighted densities
(\ref{n_alpha}) is complemented by a tensor density with Cartesian components:
  \bb
 T_{ij}(\rr)=\int\dr'\rho(\rr+\rr')\frac{r'_ir'_j}{R^2}\delta(R-|\rr'|)\,.\label{tij}
 \ee
The free-energy density is then given by
 \begin{eqnarray}
 \Phi&=&-n_0\ln(1-n_3)+\frac{n_1n_2-{\bf n}_{v1}\cdot{\bf n}_{v2}}{1-n_3}\\
 &+&\frac{3}{16\pi}\frac{{\bf n}_{v2}\cdot{\bf T}\cdot{\bf n}_{v2}-n_2n^2_{v2}-{\rm Tr}[{\bf T}^3]+n_2{\rm Tr}[{\bf T}^2]}{(1-n_3)^2}\,,\nonumber
 \end{eqnarray}
 where ${\bf T}$ is the matrix corresponding to (\ref{tij}) and where we have kept the original Rosenfeld's notation.

Minimizing Eq.~(\ref{om}) yields an Euler-Lagrange equation:
 \bb
 \frac{1}{\beta}\ln\Lambda^3\rhor+\frac{\delta{\cal F}_{\rm hs}[\rho]}{\delta\rho(\rr)}+\int\dd\rr'\rho(\rr')u_{\rm a}(|\rr-\rr'|)=\mu-V(\rr)\,.\label{el}
 \ee

An external potential $V(\rr)$ representing a corrugated wall, is constructed by considering a semi-infinite solid slab of uniform density $\rho_w$ into which an
infinite array of infinitely long narrow grooves is cut, as illustrated in Fig.~\ref{sketch}. Alternatively, the wall surface can be viewed as a planar wall onto
which an infinite array of infinitely long pillars is deposited. The depth of the grooves or the height of the pillars is $D$. The width of the pillars is $L_1$ and
the width of the grooves is $L_2$, such that $L=L_1+L_2$ is the periodicity of the potential. The wall atoms interact with the fluid particles via the Lennard-Jones
potential
 \bb
 \phi(r)=4\varepsilon_w\left[\left(\frac{\sigma}{r}\right)^{12}-\left(\frac{\sigma}{r}\right)^6\right]\,,\label{ext_field}
 \ee
such that total external potential experienced by the fluid atoms is
 \bb
 V(x,z)=V_{\pi}(z)+\sum_{n=-\infty}^\infty V_p(x+nL,z)\,,
 \ee
 where
 \bb
 V_\pi(z)=4\pi\varepsilon_w\rho_w\sigma^3\left[\frac{1}{45}\left(\frac{\sigma}{z}\right)^9-\frac{1}{6}\left(\frac{\sigma}{z}\right)^3\right]
 \label{pot_flat}
 \ee
 is the potential of the flat wall and $V_p(x,z)$ is the potential of a single pillar:
 \begin{eqnarray}
 V_p(x,z)&=&\rho_w\int_0^{L_1}\dd x'\int_{-\infty}^\infty\dd y\int_0^{D}\dd z'\nonumber\\
 &&\phi\left(\sqrt{(x-x')^2+y^2+(z-z')^2}\right)\nonumber\\
 &\equiv& V_{12}(x,z)+V_{6}(x,z)\,.
\end{eqnarray}
The $V_{12}(x,z)$ term describes the repulsive part of the wall-fluid interaction and has the form:
\begin{eqnarray}
V_{12}(x,z)&=&4\varepsilon_w\sigma^{12}\rho_w\int_{x-L_1}^{L_1}\dd x'\int_{-\infty}^\infty\dd y\int_{z-D}^{z}\dd z'
 \frac{1}{(x'^2+y^2+z'^2)^6}\nonumber\\
&=&\pi\varepsilon_w\sigma^{12}\rho_w\left[\psi_{12}(x,z)-\psi_{12}(x,z-D)\right.\nonumber\\
&&\left.-\psi_{12}(x-{L_1},z)+\psi_{12}(x-{L_1},z-D)\right]
\end{eqnarray}
where
\begin{widetext}
\bb
 \psi_{12}(x,z)=-\frac{1}{2880}\frac {128\,{x}^{16}+448\,{x}^{14}{z}^{2}+560\,{x}^{12}{z}^{4}+280\,{
x}^{10}{z}^{6}+35\,{x}^{8}{z}^{8}+280\,{x}^{6}{z}^{10}+560\,{x}^{4}{z} ^{12}+448\,{z}^{14}{x}^{2}+128\,{z}^{16}}{{z}^{9}{x}^{9} \left( {x}^{2 }+{z}^{2} \right)
^{7/2}}
 \ee
\end{widetext}

The attractive contribution from a single pillar can be conveniently written as \cite{mal_grooves}
\begin{eqnarray}
V_6(x,z)&=&-4\varepsilon_w\sigma^6\rho_w\int_{x-L_1}^{L_1}\dd x'\int_{-\infty}^\infty\dd y\int_{z-D}^{z}\dd z'\nonumber\\
 &&\frac{1}{(x'^2+y^2+z'^2)^3}\nonumber\\
  &=&\alpha_w\left[\psi_6(x,z)-\psi_6(x,z-D)\nonumber\right.\\
  &&\left.-\psi_6(x-L_1,z)+\psi_6(x-L_1,z-D)\right]\,,
\end{eqnarray}
where
\bb
 \alpha_w=-\frac{1}{3}\pi\varepsilon_w\sigma^6\rho_w
 \ee
and
 \bb
  \psi_6(x,z)=-{\frac {2\,{x}^{4}+{x}^{2}{z}^{2}+2\,{z}^{4}}{2{z}^{3}{x}^{3} \sqrt {{x}^{2}+{z}^{2}}}}\,.
 \ee

Having determined the equilibrium density profile $\rho(x,z)$ we construct the excess adsorption
 \bb
 \Gamma=\frac{1}{L}\int\int\dd x \dd z (\rho(x,z)-\rho_b)\,,\label{ads}
 \ee
where $\rho_b$ is the density of the bulk phase and  the integral is performed over one period of the system accessible volume. The excess adsorption serves as an
order parameter for determining the wetting temperature $T_w$ and is finite for $T<T_w$ and diverges for $T\geq T_w$.

 \subsection{Constrained DFT}

 In the previous paragraph a standard grand-canonical DFT was presented. In order to describe a shape of a cylindrical drop deposited on a smooth or rough surface we
also employ what we call here constrained DFT, in which case the average number of particles, rather than directly the chemical potential, is maintained fixed. Owing
to the translation symmetry of the (infinite) system along one of the directions parallel with the wall, we fix
   \bb
  \int\dd x\dd z\rho(x,z)=\frac{\langle N\rangle}{L_\parallel}\,, \label{el2}
  \ee
  where the r.h.s. of Eq.~(\ref{el2}) expresses the average number of particles per unit length of the system in the $y$-direction.

We note that the computation of the equilibrium density profiles is still performed in a grand-canonical ensemble using the same grand-canonical intrinsic free energy
functional (\ref{f}) as in the previous case, such that we minimize
 \bb
 {\cal F}[\rho]+\int\dd\rr\rhor V(\rr)\,,\label{el1}
 \ee
subject to the constraint of Eq.~(\ref{el2}). This approach should be distinguished from a canonical ensemble DFT describing closed systems, i.e. those at which the
(integer) number of particles $N$, rather than the average number of particles $\langle N\rangle$, is fixed. This canonical ensemble DFT can be performed either using
a free energy functional derived in the formalism of the canonical ensemble \cite{blum} or by linking approximately the grand canonical DFT functional with the
canonical ensemble one \cite{white1,white2,white3}. Neither is necessary for our purposes where we deal with open systems with a large number of particles (in
contrast to systems containing only few particles that are confined by a closed impenetrable cavity as considered in Refs. \cite{white1,white2,white3}). For systems
far from any phase transition, this approach is completely equivalent to the standard unconstrained DFT as the constraint imposed by Eq.~(\ref{el2}) is equivalent to
fixing a corresponding value of the chemical potential. However, a special care must be taken for systems at or near-to two-phase equilibrium. In these cases, the
constrained DFT  allows to stabilize a mixture of these two phases due to the constraint of the system but a reasonable initial conditions respecting the system
symmetry must be imposed in order to avoid artificial results.

 The constrained minimization leads to the following equation for the equilibrium density profile:
 \bb
 \rhor=\frac{N\exp\left[c^{(1)}(\rr)-\beta V(\rr)\right]}{\int\dr \exp\left[c^{(1)}(\rr)-\beta V(\rr)\right]}\,, \label{cdft}
  \ee
where
 \bb
c^{(1)}(\rr)=-\frac{\delta F_{\rm ex}[\rhor]/k_BT}{\delta\rhor}
 \ee is  the grand canonical single-particle direct correlation function.  Eq.~(\ref{cdft})
 can be solved iteratively by starting from an initial density profile $\rho(x,z)$,
such that a part of the box (presumably to be occupied by the liquid drop) has a uniform  density $\rho_l(T)$ and the rest of the box is of a uniform density
$\rho_v(T)$. Upon iterating Eq.~(\ref{cdft}) the density profile evolves toward the equilibrium  such that the entire average density of the system remains unchanged.

Both Euler-Lagrange equations (\ref{el}) and (\ref{cdft}), corresponding to DFT and cDFT, respectively, are solved using the Picard iteration for the equilibrium
profile $\rho(x,z)$ on a two-dimensional Cartesian grid with a spacing of $0.05\,\sigma$. After an appropriate transformation of the coordinate system, the
corresponding two-dimensional integrals are expressed over the interval $(-1,1)$ and approximated by the Gaussian quadrature. The value of the integrands at the
points out of the grid is evaluated by the bilinear interpolation, see Ref. \cite{mal_grooves} for more details.

\section{Results}

\begin{figure}[h]
\includegraphics[width=0.5\textwidth]{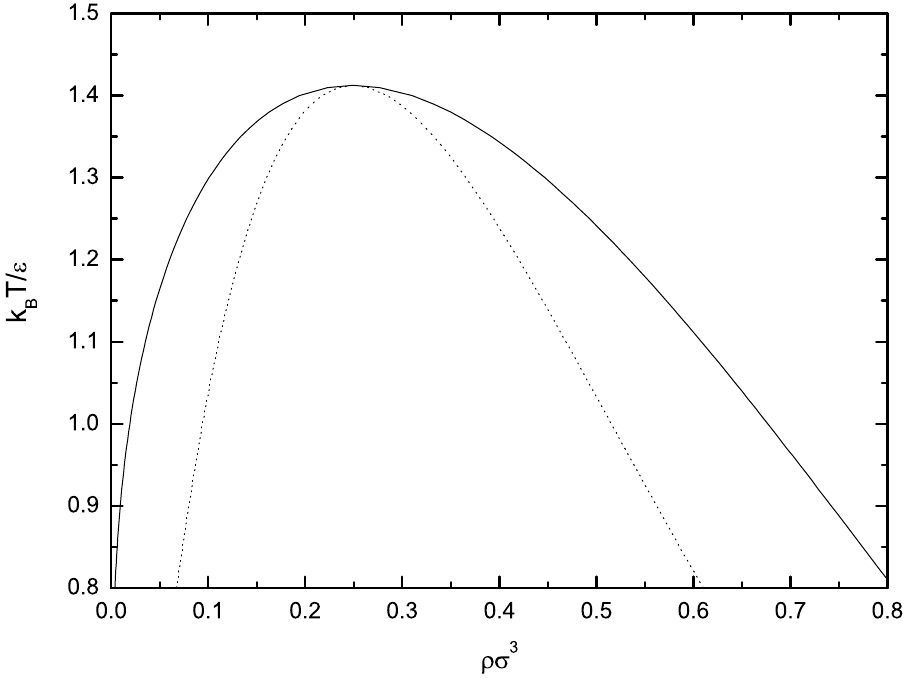}
\caption{Bulk phase diagram for a model fluid consisting of a hard-sphere repulsion (over the range of $\sigma$) and a truncated ($r_c=2.5\,\sigma$)
$-4\varepsilon\sigma^6/r^{6}$ attraction; the solid line represents  vapor-liquid coexistence (binodal), and the dotted line represents the limit-of-stability
(spinodal). } \label{fig1}
\end{figure}

\begin{figure}[h]
\includegraphics[width=0.5\textwidth]{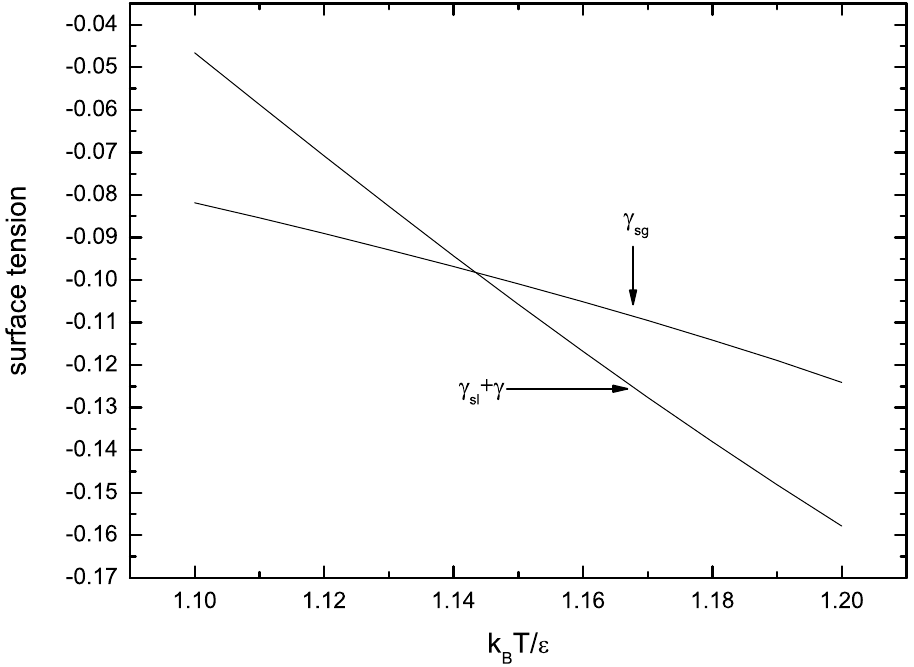}
\caption{Determination of the wetting temperature ($k_BT_w\approx1.142\,\varepsilon$) for a planar wall from the intersection of the solid-vapor surface tension
$\gamma_{sg}$ with the summed tensions $\gamma_{sl}+\gamma$; surface tensions are expressed in units of $\varepsilon/\sigma^2$. }  \label{fig2}
\end{figure}

In this section, we present numerical results of the DFT model that was described in the previous section.  Before analyzing the surface properties of the model
substrates, we present the bulk properties of our model fluid, using Eq.~(\ref{el}) for the case of zero external field. In Fig.~\ref{fig1}, we show the liquid-vapor
phase diagram in the temperature-density plane, delineating the phase boundary and the limit-of-stability. The latter is defined by the condition $\frac{\partial^2
(F/V)}{\partial\rho^2}=0$ and determines the region of the phase diagram where thermodynamic states cease to be even locally stable. The two curves terminate at a
critical point, which has coordinates $k_BT_c=1.411\,\varepsilon$ and $\rho_c=0.249\,\sigma^{-3}$, as determined by the additional condition $\frac{\partial^3
(F/V)}{\partial\rho^3}=0$.

\subsection{Wetting on a planar wall}

\begin{figure}[h]
\includegraphics[width=0.5\textwidth]{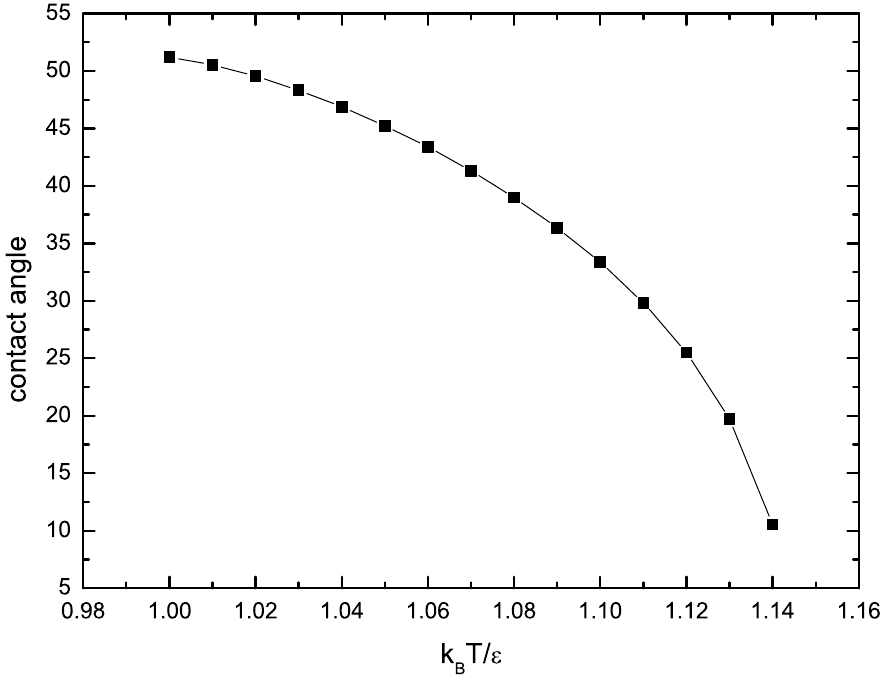}
\caption{Temperature dependence of the contact angle (in degrees) of a sessile drop resting on an ideally planar wall.}  \label{fig3}
\end{figure}

To begin, we examine the wetting properties of a ``reference system'' of a planar wall with the potential given by Eq.~(\ref{pot_flat}). This system corresponds to a
one-dimensional problem in which the equilibrium density profile varies in a single Cartesian coordinate, $\rhor=\rho(z)$. Nevertheless, for numerical consistency,
the system is treated in the same manner as the models of the corrugated walls, in which the external potential, and thus the equilibrium density profile, is
two-dimensional. The comparison between the results from one-dimensional and two-dimensional treatments of the system revealed a very good agreement which provides
good verification of our numerical methods.

Throughout the study the wall strength is fixed at $\varepsilon_w=\varepsilon$. As shown below, the wetting temperature of the flat wall is well below the bulk
critical temperature for this value of the wall parameter, which avoids potential complications from critical fluctuations near $T_c$ in view of the mean-field nature
of DFT. On the other hand, this value of $\varepsilon_w$ is sufficiently small for the system to exhibit layering transitions, the study of which is beyond the scope
of this work.

Perhaps the most straightforward approach to determine the wetting temperature $T_w$ at which the thickness of the adsorbed liquid film diverges (and the contact
angle of the sessile drop vanishes) is to use Young's equation (\ref{young}), which transforms to Antonoff's rule $\gamma_{\rm sv}(T_w)=\gamma_{\rm
sl}(T_w)+\gamma(T_w)$ exactly at $T_w$. The surface tensions $\gamma_{\rm sv}$ and $\gamma_{\rm sl}$ are obtained by minimizing the grand potential functional given
in Eq.~(\ref{om}) subject to the boundary conditions $\rho(x,z_M)=\rho_v$ and $\rho(x,z_M)=\rho_l$, $\forall x$, respectively, where $\rho_v$ and $\rho_l$ are the
coexisting densities and $z_M$ is the size of the system in the $z$-dimension. Of course, for a planar wall, the density profile $\rho(x,z)$ does not depend on the
$x$-coordinate (along the wall); additional periodic boundary conditions $\rho(0,z)=\rho(x_M,z)$, $\forall z$, are imposed for the corrugated walls that are
considered in the following paragraph, with $x_M$ being the size of the system in the $x$-dimension. The liquid-vapor surface tension $\gamma$ is determined
independently by equilibrating a system filled with the two coexisting fluid phases in the absence of an external field. Here, we verify again that both numerical DFT
treatments produce consistent results.

\begin{figure*}
\includegraphics[width=0.23\textwidth]{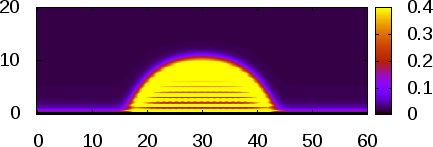}\;\;\includegraphics[width=0.23\textwidth]{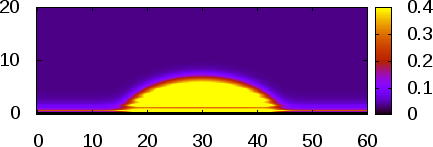}\;\;\includegraphics[width=0.23\textwidth]{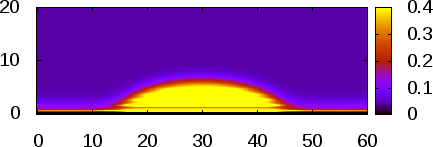}\;\; \includegraphics[width=0.23\textwidth]{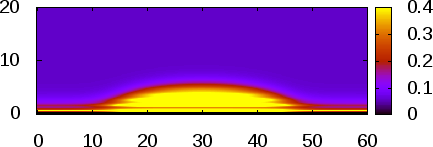}
\includegraphics[width=0.23\textwidth]{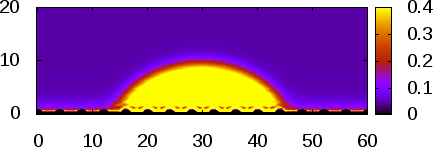}\;\;\includegraphics[width=0.23\textwidth]{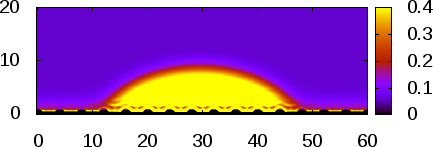}\;\;\includegraphics[width=0.23\textwidth]{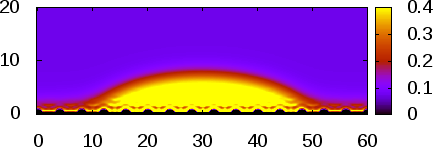}\;\;\includegraphics[width=0.23\textwidth]{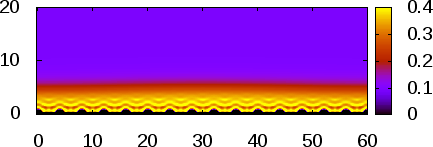}
\includegraphics[width=0.23\textwidth]{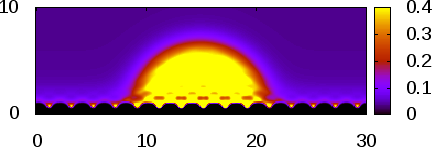}\;\;\includegraphics[width=0.23\textwidth]{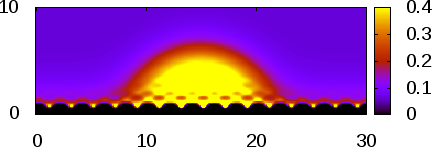}\;\;\includegraphics[width=0.23\textwidth]{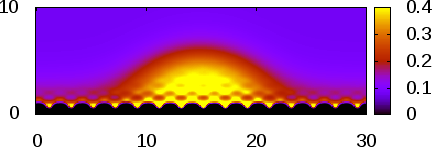}\;\;\includegraphics[width=0.23\textwidth]{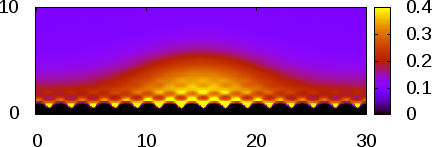}
\includegraphics[width=0.23\textwidth]{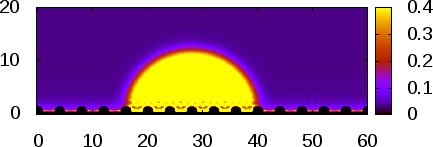}\;\;\includegraphics[width=0.23\textwidth]{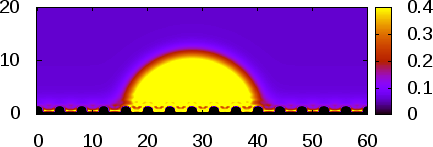}\;\;\includegraphics[width=0.23\textwidth]{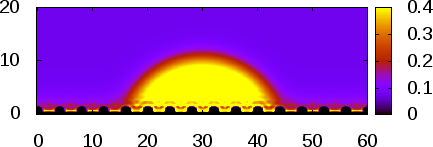}\;\;\includegraphics[width=0.23\textwidth]{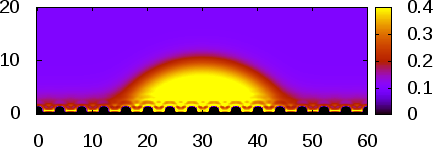}
\includegraphics[width=0.23\textwidth]{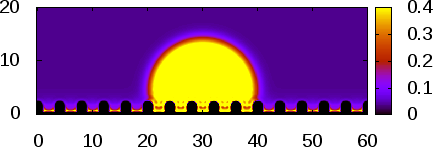}\;\;\includegraphics[width=0.23\textwidth]{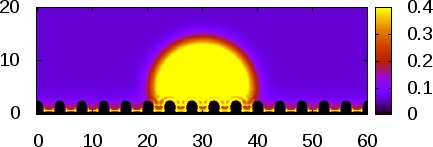}\;\;\includegraphics[width=0.23\textwidth]{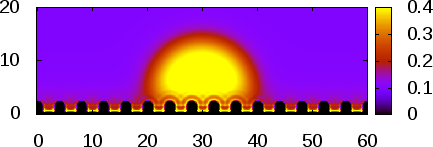}\;\;\includegraphics[width=0.23\textwidth]{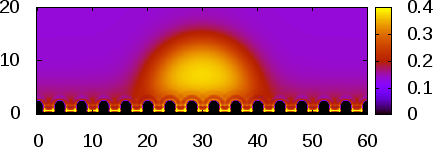}
\includegraphics[width=0.23\textwidth]{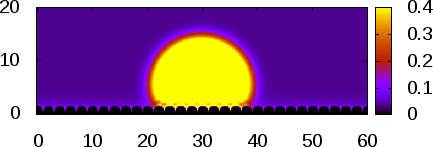}\;\;\includegraphics[width=0.23\textwidth]{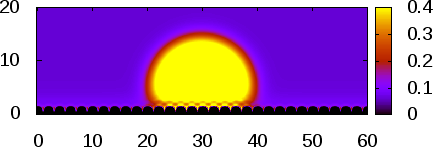}\;\;\includegraphics[width=0.23\textwidth]{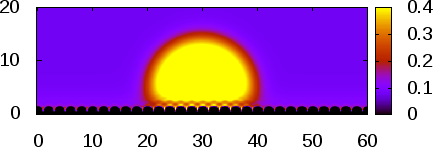}\;\;\includegraphics[width=0.23\textwidth]{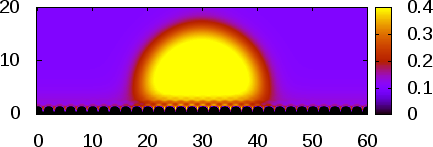}
\includegraphics[width=0.23\textwidth]{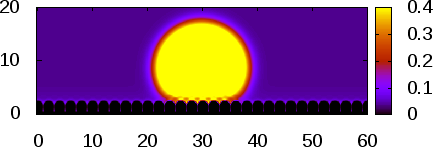}\;\;\includegraphics[width=0.23\textwidth]{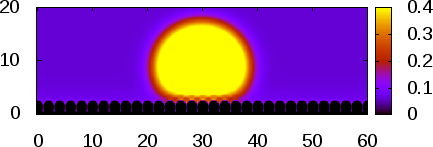}\;\; \includegraphics[width=0.23\textwidth]{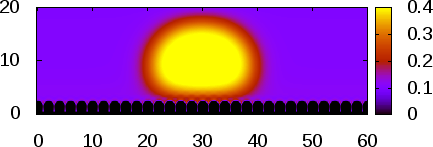}\;\;\includegraphics[width=0.23\textwidth]{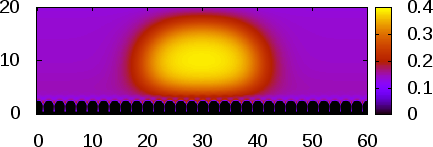}
\caption{ Two-dimensional density profiles of a sessile drop on  walls of different corrugations. {\bf First row}: Case 1 (smooth wall); the profiles correspond to
the temperatures (from left to right): $k_BT/\varepsilon=1$, $k_BT/\varepsilon=1.1$, $k_BT/\varepsilon=1.12$ and $k_BT/\varepsilon=1.14$. {\bf Second row}: Case 2;
the profiles correspond to the temperatures (from left to right): $k_BT/\varepsilon=1.1$, $k_BT/\varepsilon=1.15$, $k_BT/\varepsilon=1.2$ and $k_BT/\varepsilon=1.25$.
{\bf Third row}: Case 3; the profiles correspond to the temperatures (from left to right): $k_BT/\varepsilon=1.1$, $k_BT/\varepsilon=1.2$, $k_BT/\varepsilon=1.25$ and
$k_BT/\varepsilon=1.28$. {\bf Fourth row}:  Case 4; the profiles correspond to the temperatures (from left to right): $k_BT/\varepsilon=1.1$, $k_BT/\varepsilon=1.15$,
$k_BT/\varepsilon=1.18$ and $k_BT/\varepsilon=1.2$. {\bf Fifth row}: Case 5; the profiles correspond to the temperatures (from left to right): $k_BT/\varepsilon=1.1$,
$k_BT/\varepsilon=1.2$, $k_BT/\varepsilon=1.3$ and $k_BT/\varepsilon=1.35$. {\bf Sixth row}: Case 6; the profiles correspond to the temperatures (from left to right):
$k_BT/\varepsilon=1.1$, $k_BT/\varepsilon=1.15$, $k_BT/\varepsilon=1.18$ and $k_BT/\varepsilon=1.2$. {\bf Seventh row}: Case 7; the profiles correspond to the
temperatures (from left to right): $k_BT/\varepsilon=1.1$, $k_BT/\varepsilon=1.2$, $k_BT/\varepsilon=1.3$ and $k_BT/\varepsilon=1.35$. In all cases, the equilibrium
density distributions correspond to systems of dimensions  $60\,\sigma\times20\,\sigma$ and are expressed in units of $\sigma$ and $\varepsilon$. Each case
corresponds to a substrate model according to Table 1.} \label{profiles}
\end{figure*}

In Fig.~ \ref{fig2}, we show the temperature dependence  of the solid-vapor surface tension $\gamma_{sv}$ and the summed tensions $\gamma_{sl}+\gamma$. The
intersection of the two curves determines the wetting temperature $k_BT_w\approx1.142\,\varepsilon$.
We further use Young's equation (\ref{young}) to determine the temperature dependence of the contact angle $\theta(T)$, as shown in Fig.~\ref{fig3}. These results can
be compared with those obtained from cDFT that is given by Eq.~(\ref{cdft}). The resulting two-dimensional density profiles for several representative temperatures
are displayed in Fig.~\ref{profiles} (first row).

\begin{figure}
\includegraphics[width=0.5\textwidth]{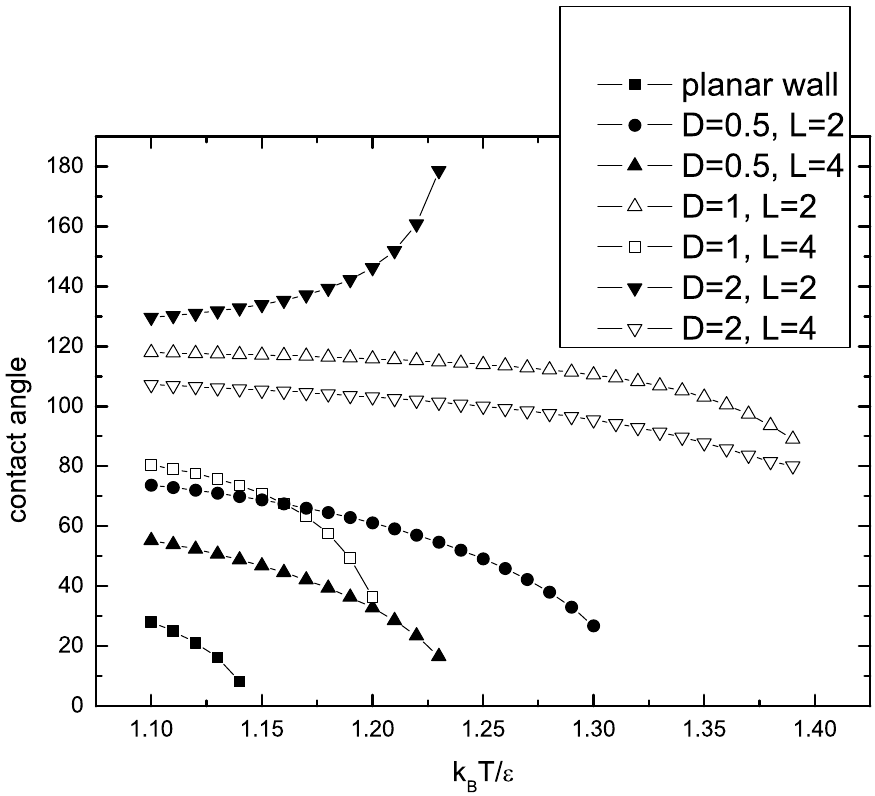}
\caption{Comparison of the temperature dependence of the contact angle  (in degrees) for walls with different types of corrugation.} \label{theta_rough}
\end{figure}

\subsection{Wetting properties of corrugated walls}

\begin{table}
\begin{tabular}{cccc}
\hline
 case No.& model parameters & $r$ & contact angle\\
\hline
1& planar wall &1& 28\\
2& $D=0.5\sigma,\,L=4\sigma$ & 1.25 & 58\\
3& $D=0.5\sigma,\,L=2\sigma$ & 1.5  & 77\\
4& $D=1.0\sigma,\,L=4\sigma$ & 1.5  & 80\\
5& $D=2.0\sigma,\,L=4\sigma$ &2 &   108\\
6& $D=1.0\sigma,\,L=2\sigma$ & 2& 116\\
7& $D=2.0\sigma,\,L=2\sigma$ & 3& 129\\
\hline
\end{tabular}
\caption{Contact angles at a temperature corresponding to $k_BT=1.1\,\varepsilon$ for walls with different types of corrugation; the corrugation is formed by pillars
of depth $D$, width $L_1=0.5\,\sigma$ and a periodicity $L$; in the second column, the ``surface roughness'' of each of the models is estimated as $r\approx1+2D/L$;
in the third column, the contact angle of the sessile drop is expressed in degrees. } \label{tab}
\end{table}



\begin{figure}
\includegraphics[width=0.4\textwidth]{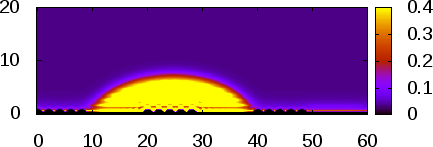}
\includegraphics[width=0.4\textwidth]{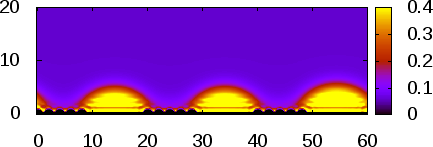}
\includegraphics[width=0.4\textwidth]{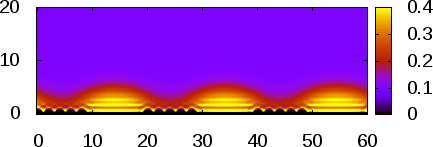}
\caption{Two-dimensional density profiles on a wall where smooth and rough parts of the surface alternate. Each rough part is made out of five steps with corrugation
parameters $D=0.5\,\sigma$ and $L=2\,\sigma$. The equilibrium profiles are obtained from a constrained DFT and correspond to the temperatures (from  top  to
 bottom): $k_BT/\varepsilon=1.1$, $k_BT/\varepsilon=1.2$ and $k_BT/\varepsilon=1.3$.  The equilibrium
density distributions correspond to systems of dimensions  $60\,\sigma\times20\,\sigma$ and are expressed in units of $\sigma$ and $\varepsilon$. } \label{mix_cdft}
\end{figure}

\begin{figure}
\includegraphics[width=0.4\textwidth]{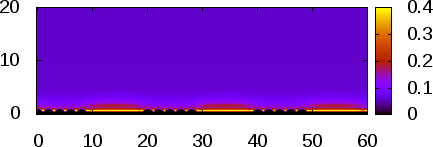}
\includegraphics[width=0.4\textwidth]{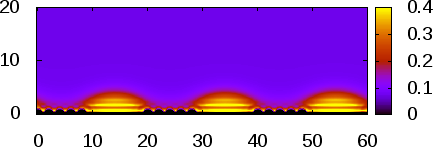}
\includegraphics[width=0.4\textwidth]{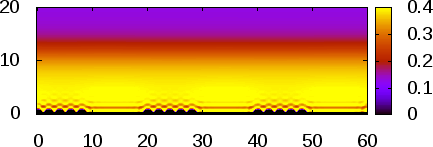}
\caption{Two-dimensional density profiles on a wall where smooth and rough parts of the surface alternate. Each rough part is made out of five steps with corrugation
parameters $D=0.5\,\sigma$ and $L=2\,\sigma$. The equilibrium profiles are obtained from a grand canonical ensemble DFT and correspond to the temperatures (from the
top to the bottom): $k_BT/\varepsilon=1.1$, $k_BT/\varepsilon=1.2$ and $k_BT/\varepsilon=1.35$.  These results should be compared with figure \ref{theta_rough}.
 The equilibrium density distributions correspond to systems of dimensions  $60\,\sigma\times20\,\sigma$ and are expressed in units of $\sigma$ and
$\varepsilon$.} \label{mix_dft}
\end{figure}

We now turn our attention to the wetting properties of corrugated walls. To this end, we fix the width of each pillar at $L_1=0.5\,\sigma$ and consider three
different pillar heights ($D=0.5\,\sigma$, $D=\sigma$ and $D=2\,\sigma$) and two periodicities ($L=2\,\sigma$ and $L=4\,\sigma$). Somewhat naively, we can assign a
roughness parameter value to each of the substrate models as follows:
 \bb
 r\equiv 1+2D/L\,, \label{roughness}
 \ee
and search for the correlation between $r$ and the surface wettability.  Note that this simple definition of the  roughness parameter is somewhat arbitrary and that
alternative possibilities how the define surface roughness are available \cite{whitehouse, gadel}. First, we fix the temperature at $k_BT=1.1\,\varepsilon$
($T=0.78\,T_c$), which is below the wetting temperature for the planar wall. From Table 1, where we show the contact angle for each substrate, we find that: i)
somewhat surprisingly, the contact angles of different models with identical $r$ almost coincide, which supports the definition (\ref{roughness}), and ii) the contact
angle increases with $r$. For relatively high $r$, the increase in the apparent angle is sufficiently pronounced, such that the hydrophilic surface ($\theta<\pi/2$)
becomes hydrophobic ($\theta^*>\pi/2$). Note that these observations are in direct contradiction with Wenzel's law given by Eq.~(\ref{wenzel}).

Next, we consider a temperature dependence of the contact angle. The most important conclusion that can be drawn from the results shown in Fig.~\ref{theta_rough} is
that corrugation always decreases the hydrophilicity of the wall. More specifically, for relatively low corrugation (small $r$), the surface roughness shifts the
wetting temperature to significantly higher values than of the planar wall. For more strongly corrugated walls, the role of the roughness is even more dramatic as the
walls become non-wetting. The models with the roughness parameter $r=2$ become hydrophobic over almost the entire range of temperatures apart from the immediate
vicinity of the critical temperature where $\theta\approx90\degree$. Most notably, when the surface effects become sufficiently strong (corresponding to the model
with the highest roughness parameter), the contact angle {\emph{increases}} with the temperature, which ultimately leads to {\emph{drying}} at a temperature
corresponding to $k_BT_d\approx1.24\,\varepsilon$.

These results can be compared with representative two-dimensional density profiles that are obtained using  cDFT (see Fig.~\ref{profiles}) to produce the states of
the sessile drop for the same substrates as considered in Fig.~\ref{theta_rough}. Here, we present the equilibrium profiles corresponding to relatively large systems
with a lateral and a normal size of $60\,\sigma$ and $20\,\sigma$, respectively. We have also tested our results for different system sizes, different total amounts
of fluid molecules and different initial states.  Since for our model the wetting transition is first order, there is a free-energy barrier between partial and
complete wetting states. In order to avoid a situation where the system is trapped in a local minimum of the free energy,  we have started from two distinct initial
configurations: one corresponding to a partial wetting ($\theta>0$) state and one corresponding to a complete wetting ($\theta=0$) state, and considered the one with
a lower value of the free energy. We observe that the hydrophilic nature of the substrates decreases with the roughness parameter which is consistent with the results
that are obtained using the grand canonical DFT. In particular, while for $r=1.25$--$1.5$  the liquid tends to spread over the solid surface as the temperature
increases and for $r=2$  the contact angle is high and nearly constant, there is an increase in the contact angle of the liquid drop for $r=3$ as the temperature
increases. From a macroscopic perspective, the case shown in the lowest row of Fig.~\ref{profiles} provides a so called Cassie-Baxter (or fakir) state, where the drop
sits on top of the surface with the vapor trapped underneath,  even though the material of the wall is hydrophilic; this situation occurs for the temperatures
$k_BT/\varepsilon=1.1$ and $k_BT/\varepsilon=1.2$. At the higher temperatures the liquid drop already ``levitates'' above the wall, in agreement with the results
shown in Fig.~\ref{theta_rough}.

Finally, we consider a substrate with a surface of alternating smooth and corrugated regions of equal areas. For the corrugated section, we use  a model with
$D=0.5\,\sigma$, $L_1=0.5\,\sigma$ and $L=2\,\sigma$ (case 3). In Fig.~\ref{mix_cdft}, we display the equilibrium density profiles that are obtained by cDFT. At
higher temperatures, the liquid separates into cylindrically shaped structures of identical cross sections above the smooth parts of the surface, while the corrugated
parts remain dry. However, when the temperature is lowered sufficiently, the liquid undergoes a morphological transition and forms a single bulge. This transition can
be explained in terms of the competition between the dewetting of the rough section of the wall by the liquid and the energetic cost of creating a liquid-vapor
interface, which is high at low temperatures.

These results are complemented by a standard grand canonical DFT calculation, with an ambient phase of saturated vapor, see Fig.~\ref{mix_dft}. For a temperature
below the wetting temperature of the corresponding planar wall $T_w$, only a microscopic layer of liquid forms at the smooth sections and impregnates the tiny grooves
in the rough parts. Above $T_w$ but below the wetting temperature of the corrugated section $T_w^r$,  cylindrical segments of liquid form at the smooth sections of
the surface, which corresponds to the high-temperature state shown in Fig.~\ref{mix_cdft}. Finally, the entire surface is completely  wet above $T_w^r$. Note that
although the liquid-vapor interface is flat, and thus does not reflect the shape of the wall, the lateral inhomogeneity in the liquid structure remains fairly
pronounced near the wall.

\section{Summary of the results}

\begin{figure*}
\includegraphics[width=0.5\textwidth]{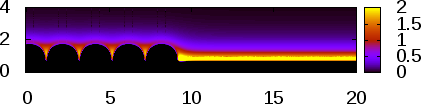}
\caption{Modulus of the external potential  (expressed in units of $\varepsilon$) of a substrate consisting of rough ($D=\sigma$, $L_1=0.5\,\sigma$) and smooth
sections in the $x$-$z$ plane; the plot shows that the presence of the pillar effectively weakens the  net potential of the wall except inside the grooves.  The
coordinates correspond to units of $\sigma$.} \label{pot}
\end{figure*}

In this paper, we have studied the wetting properties of microscopically rough surfaces of hydrophilic material. The methods used in the study and the main results
are summarized below:

\begin{itemize}

\item We have applied a density functional theory that is based on the Tarazona tensorial version of FMT,  which
is most likely the most accurate microscopic approach available for non-uniform fluids.  This FMT-DFT avoids the spurious divergencies produced by the original
Rosenfeld FMT version, satisfies exact statistical mechanical sum rules and provides correct limits in reduced dimensions.

\item We have started with a detailed description of a ``reference system'' of an ideally planar wall interacting with the fluid particles via long-ranged
dispersion forces. The wall exhibits a first-order wetting transition (well below the critical temperature, $T_w/T_c=0.8$)  that can be characterized as the
temperature at which a macroscopically thick liquid layer forms at the wall or, equivalently, as the temperature at which the contact angle of a sessile drop
vanishes. We have used both fixed-$\mu$ and fixed-$\langle N\rangle$ DFT to determine $T_w$ and the temperature dependence of the contact angle.

\item   We have found that microscopic surface roughness always \emph{deteriorates} the wall wettability.
This result is in direct contradiction with the classical Wenzel's law, which predicts that the wetting properties of the wall are always \emph{amplified} by surface
corrugation, such that the hydrophilicity of surfaces increases with surface roughening. We discuss this point below.

\item For a liquid deposited on substrates with alternating flat and rough sections  we observed two possible morphologies depending
on the temperature. At  high temperatures, the liquid tends to wet the planar sections of the wall and forms periodic cylindrical segments within which the rough
sections remain dry. However, as the temperature decreases, a configuration with only one liquid bulge is favored, in view of the high energetic cost of creating a
liquid-vapor interface.

\end{itemize}

\section{Concluding remarks}

 The main conclusion  of this study is that DFT results of wetting properties of microscopically rough surfaces contradict macroscopic predictions, such as those
implied by Wenzel's law. To understand this point we briefly revisit the arguments leading to Eq.~(\ref{wenzel}): assuming the liquid-wall interface beneath a drop
deposited on a rough surface follows the wall corrugation, we minimize the surface energy for a virtual displacement $\dd x$ of the contact line
%
%
 \bb
 \dd E=r\left(\gamma_{sl}-\gamma_{sv}\right)\dd x+\gamma\dd x\cos\theta^*\,.
 \ee
Within this picture the effect of the surface roughness  only manifests as an increase in the substrate area by a factor of $r$, whereas the other aspects of surface
corrugation are ignored. To interpret our DFT results, we now present the meso- and microscopic implications of the non-planar wall geometry. We consider nearly flat
wall with ideal planarity that is perturbed by adding (or removing) a microscopically small portion of the material at some point on (from) its surface. The
perturbation can be represented by a single pillar (or a single groove) that is deposited (etched) on (into) the wall, although the specific shape of the barrier
(well) is not crucial. More importantly, the mean height of the liquid-vapor interface is now a function of the horizontal position $\ell=\ell(x)$. To be specific, we
may consider a complete wetting regime, such that the pressure is slightly below its saturation value.  If the planar wall is perturbed, the liquid-vapor interface
must bend around the barrier, which costs an additional local bending energy per unit area, $\delta E\approx\gamma/2(\nabla\ell)^2$; thus the wall barrier also acts
as an energy barrier. This effect is also responsible for the binding of the local interface near the edge of an apex-shaped substrate \cite{parry_apex, mal_apex} and
for the finite thickness of a wetting layer on a spherical wall \cite{bieker, stewart, nold}, even in the limit of bulk coexistence.

Let us now consider a semi-infinite rough surface ($D=\sigma$, $L_1=0.5\,\sigma$) and compare the magnitude of the wall potential with that corresponding to a
semi-infinite smooth surface, as shown in Fig.~\ref{pot}. We can immediately draw two conclusions: i) Unlike the potential for the smooth section, the potential above
the rough surface exhibits lateral inhomogeneity, although weakly. The shape of the adsorbed film tends to follow the geometry of the external field; therefore,
adsorption at the rough surface is more energetically expensive because of the aforementioned cost of the surface tension. ii) More importantly, the excluded volume
effects from the presence of the pillars produce a region that is inaccessible to the fluid molecules, that have a significantly larger volume than the actual volume
of the pillars. Note that as an effect the rectangular pillars appear rounded. Consequently, the external field above the pillars is markedly lowered relative to the
smooth surface, which explicitly demonstrates why surface roughness hinders adsorption. On the other hand, there is a fairly strong field within the accessible volume
between the pillars, i.e., inside the grooves. Therefore, it is much easier for the liquid to fill the grooves than wet the structured surfaces, which is completely
in agreement with the macroscopic arguments \cite{quere}.

Finally, it is important to stress out that on a \emph{macroscopic} scale the wall geometry does indeed promote fluid adsorption. A simple example of this is a
wedge-like cavity, the geometry of which enhances the adsorption and decreases the effective contact angle. In a wedge with a tilt angle $\alpha$ the roughness
parameter is $r=\sec\alpha$. For $\theta<\alpha$ the wedge is completely filled ($\theta^*=0$) even though the side walls are partially wet \cite{rejmer, wood,
binder03, our_prl} and thus the wedges induce the filling transition at which a macroscopic amount of liquid occurs at a temperature $T_f<T_w$. This is in a complete
qualitative accord with Wenzel's law. However, such a scenario is only conceivable if the wedge-like structure has macroscopic dimensions which is certainly not the
case of our corrugation model. From this it follows that there must exist a \emph{crossover} between a microscopic scale, where the surface structure suppresses fluid
adsorption, and a macroscopic scale, where the adsorption is enhanced by the wall geometry.  This, we believe, is an important point in terms of the modern trends in
engineering branches to apply phenomenological theories on nanoscopically small systems (so called nanothermodynamics). The danger of these attempts has been clearly
exemplified here.

 We stress here once again that the choice to represent the macroscopic treatment of rough surfaces by Eq.~(\ref{wenzel}) was rather arbitrary and that Wenzel's law
plays no prominent role within the class of phenomenological theories. The only specific feature of Wenzel's law is that it predicts that the surface roughness lowers
the wetting temperature, i.e. that $\theta^*$ may vanish even though $\theta>0$. Note that this prediction is not supported by the experiments on textured surfaces
\cite{kao} which revealed that the wetting (or drying) temperature is unaffected by the surface roughness. This is because, in the hydrophilic case, the Wenzel's
regime fails to describe a propagation of liquid film in the solid grooves (so-called hemi-wicking \cite{quere, bico}). If this second phenomenon is taken into
account it follows that $\theta^*=0$ only if $\theta=0$ but even in this regime the surface wettability is improved ($\theta>\theta^*>0$). This still contradicts our
microscopic results, however.



 In this work, we dealt with a class of very simple models of geometrically nonuniform surfaces and of course further investigations in this direction are needed.
One cannot exclude that the picture of wetting on more realistic model surfaces is more complex than that made in this work. This study can be most directly extended
in several ways. Fixed molecular parameters have been considered throughout this study and only the wall geometry and temperature were allow to vary. In particular,
the amplitudes of the wall-fluid and fluid-fluid interactions were assumed to be identical. It would be interesting to investigate whether our conclusions remain
unchanged if the wall potential strength is allowed to vary. One can speculate, that for $\varepsilon_w<\varepsilon$, i.e., a weak substrate, the liquid would
penetrate into the wall pockets effectively smoothing out the surface, which, as a result, would become a stronger adsorbent than the solid planar wall. On the other
hand, for an appreciably stronger wall potential than that used in our model,  a sequence of layering transitions may occur; it would then be interesting to determine
whether the interfaces between the neighboring layers are planar or whether their structure mirrors the wall geometry. Extending the fluid model to non-spherical
bodies, such as model polymers or liquid crystals, would be challenging and computationally demanding, and has only been attempted rather recently \cite{patricio}.
Finally, it would be interesting, especially in view of the results presented in Figs.~\ref{mix_dft} and \ref{mix_cdft}, to investigate the dynamical aspects of the
model and to track the system evolution from a particular initial state toward equilibrium.

\begin{acknowledgments}

\hspace*{0.01cm}

\noindent  I acknowledge the financial support from the Czech Science Foundation, Grant No. 13-09914S.
\end{acknowledgments}


\end{document}